\documentclass[twocolumn,10pt]{article}

\usepackage[a4paper,margin=2.0cm]{geometry}
\usepackage{amsmath,amssymb}
\usepackage{graphicx}
\usepackage{xcolor}
\usepackage{tikz}
\usepackage{pgfplots}
\pgfplotsset{compat=1.16}
\usetikzlibrary{arrows.meta,positioning,decorations.pathmorphing,shapes.geometric}
\usepackage{booktabs}
\usepackage{threeparttable}
\usepackage{makecell}
\usepackage{caption}
\usepackage[colorlinks=true,allcolors=blue]{hyperref}

\usepackage{titlesec}
\renewcommand{\thesection}{\Roman{section}}
\renewcommand{\thesubsection}{\Alph{subsection}}
\titleformat{\section}[block]{\centering\bfseries\scshape}{\thesection.}{0.6em}{}
\titleformat{\subsection}[block]{\centering\bfseries}{\thesubsection.}{0.6em}{}
\titlespacing*{\section}{0pt}{1.2\baselineskip}{0.5\baselineskip}

\newcommand{\epem}{$e^{+}e^{-}$}
\newcommand{\Cthree}{C$^{3}$}
\newcommand{\sqrts}{$\sqrt{s}$}

\begin{document}

\twocolumn[
  \begin{@twocolumnfalse}
  \begin{center}
    {\Large\bfseries Gravity-Aligned Vertical Electron-Positron Linear Collider\par}
    \vspace{0.8em}
    {\large Tetsuo Abe\par}
    {\texttt{<tetsuo.abe@kek.jp>}\par}
    \vspace{0.3em}
    {\itshape High Energy Accelerator Research Organization (KEK), Japan\par}
    \vspace{0.2em}
    {(Dated: July 24, 2026)\par}
    \vspace{1.0em}
  \end{center}
  \begin{quotation}
  \noindent\small\textbf{Abstract}---%
  A gravity-aligned vertical electron-positron linear collider is
  proposed as a concept to break the cost-and-scale impasse that confronts
  conventional horizontal Higgs factories. The accelerating axis is oriented
  along the local gravitational field inside a single deep vertical shaft.
  Positrons are injected downward from the surface while electrons are
  accelerated upward from the bottom, the two beams meeting at an interaction
  point located deep underground. Targeting a staged center-of-mass energy ($\sqrt{s}$) in
  the Higgs factory range ($250$--$500$~GeV), 
  this concept assumes high-gradient, high-efficiency, normal-conducting accelerating structures operated at cryogenic temperature
  together with a two-beam power-delivery scheme. It can be designed to reuse
  the existing KEK/SuperKEKB accelerator complex at the KEK Tsukuba Campus. 
  We discuss the advantages of the vertical orientation, namely a small 
  surface footprint, intrinsic azimuthal symmetry of the gravity load, 
  deep vibration isolation of the collision point, 
  gravity-assisted cryogenics, 
  a natural vertical alignment reference,
  a cosmically quiet deep interaction point,
  an opportunity for on-site electricity self-sufficiency, and 
  a unique opportunity in social/industrial evaluation, as 
  well as the principal disadvantages and challenges, including the required
  accelerating gradient and the construction of a kilometers-deep shaft. 
  Using only publicly available information, we argue that excavating a shaft with a depth of
  $\sim$3~km is feasible at a geologically favorable site.
  This paper is deliberately concept-level and is intended to motivate technological
  innovation, feasibility studies, and the exploration of possible new channels
  for Beyond-Standard-Model searches through vertical collisions.
  \vspace{1.2em}
  \end{quotation}
  \end{@twocolumnfalse}
]

\section{Introduction\label{sec:intro}}
The electron-positron (\epem) Higgs factory is the most
important challenge in high-energy physics in the search for
physics Beyond the Standard Model (BSM). Several projects around the world aim
to construct such a factory, including the International Linear Collider (ILC)~\cite{ilc1,ilc2,ilc3,ilc4,ilc5,ilc6},
the Compact Linear Collider (CLIC)~\cite{clic1,clic2,clic3,clic4}, the Cool
Copper Collider (\Cthree)~\cite{c3a,c3b}, the Future Circular Collider in
its \epem\ mode (FCC-ee)~\cite{fcc1,fcc2,fcc3,fcc4}, and
 the Circular Electron Positron Collider (CEPC)~\cite{cepc1,cepc2,cepc3}.
Among them, the ILC is the
most mature project and thus closest to becoming a reality. Nevertheless, the
ILC has been negatively reviewed by the government of the leading host country~\cite{ilcreview}, and 
its path toward realization remains uncertain.
The main factor behind this situation is the scale of the budget required for such a
large project within an academic research field. There are two approaches to
overcoming this challenge: (i) technological innovation and (ii) an expansion of the
scope of funding sources, both of which have been pursued in the ILC project, but with the results described above. 
Technological innovation usually takes a great deal of time and significantly
expanding the scope of funding requires changing the framework of the project
itself.

Here, we propose a new concept intended to break through this
impasse, namely a gravity-aligned vertical \epem\ linear collider (hereafter referred to simply as a vertical linear collider). This paper focuses
on the concept rather than on detailed technological descriptions or in-depth
discussions. It is meant to inspire and motivate the high-energy,
particle-physics, and astrophysics communities to pursue technological
innovation and feasibility studies, and to open new channels for BSM searches through vertical collisions.

\section{Concept\label{sec:Concept}}
\subsection{Overview\label{sec:Concept:Overview}}
\begin{figure}[t]
\centering
\begin{tikzpicture}[font=\footnotesize,>=Latex]
  \fill[brown!18] (-2.2,0) rectangle (2.2,-7.2);
  \draw[thick] (-2.2,0)--(2.2,0);
  \foreach \x in {-2.0,-1.5,...,2.0}{\draw[brown!55] (\x,0)--(\x-0.18,0.18);}
  \node[anchor=south west] at (-2.2,0.07){\scriptsize Ground level};
  \draw[gray!50,thick] (-0.55,0)--(-0.55,-7.0);
  \draw[gray!50,thick] (0.55,0)--(0.55,-7.0);
  \draw[->,very thick,black!70] (-1.7,-1.0)--(-1.7,-2.6) node[midway,left]{$\vec{g}$};
  \draw[->,very thick,red] (0.,-0.2)--(0.,-3.5);
  \node[red] at (0.3,-0.55){\scriptsize $e^{+}$};
  \draw[->,very thick,blue] (0.,-6.7)--(0.,-3.55);
  \node[blue] at (0.3,-6.35){\scriptsize $e^{-}$};
  \fill[yellow!85!orange] (0,-3.5) circle (2.2pt);
  \draw[thick,green!45!black] (-0.42,-3.5) rectangle (0.42,-3.5+0.0);
  \draw[thick,green!45!black] (-0.40,-3.92) rectangle (0.40,-3.08);
  \node[green!45!black,anchor=west] at (0.55,-3.5){\scriptsize IP \& detector};
  \draw[<->] (-1.15,0)--(-1.15,-7.0) node[midway,fill=white,rotate=90]{\scriptsize Depth: $\sim 3$~km};
\end{tikzpicture}
\caption{Schematic diagram of gravity-aligned vertical \epem\ linear collider.
The beam axis is parallel to gravity $\vec{g}$; positrons are accelerated
downward and electrons upward, colliding at a deep IP.}
\label{fig:concept}
\end{figure}
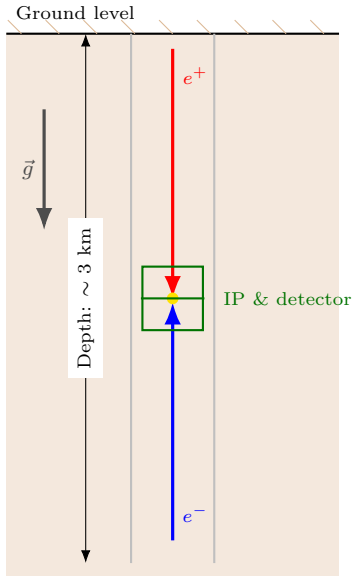

The central idea is to rotate the linear collider so that its beam axis
 is parallel to the local gravity, and to house the
machine in a single deep vertical shaft rather than in a long horizontal tunnel,
 as shown in Fig.~\ref{fig:concept}.
Positrons are injected from the surface and accelerated downward,
 while electrons are accelerated upward from the bottom of the shaft;
 the two beams collide at an interaction point (IP) located deep underground,
  where the detector is installed.
A horizontal footprint that would otherwise extend over many kilometers is thereby folded into a compact area at
 the surface, with the accelerator length converted into depth.

Because the dominant motivation of an \epem\ Higgs factory is precision physics
at a center-of-mass energy of \sqrts\ $=250$--$500$~GeV, each beam must reach an
energy on the order of $125$--$250$~GeV. Confining the machine to a vertical extent of
a few kilometers therefore requires high accelerating gradients (Sec.~\ref{sec:Concept}\,\ref{sec:Concept:Tech}).
By increasing the accelerating gradient,
 this concept is agnostic to the energy staging,
  which may follow the established Higgs factory program: \sqrts\ near $250$~GeV for the $ZH$ recoil and
Higgs couplings, near the $t\bar t$ threshold for the top-quark mass, and up to
$500$~GeV for the top Yukawa coupling and the trilinear Higgs self-coupling.

\subsection{Necessary Technologies\label{sec:Concept:Tech}}
The realization of this concept rests on three technology pillars: (i) high-gradient, 
 high-efficiency accelerating structure; (ii) ultra-low-emittance high-current electron gun; and
(iii) deep excavation engineering. Items (i) and (ii) are shared with all compact
linear collider efforts, whereas (iii) is specific to the vertical geometry.

\paragraph{High-gradient, high-efficiency accelerating structure.}
With the machine length limited to a few kilometers, an average accelerating
gradient of much higher than $100$~MV/m is required.
Superconducting radio-frequency (RF) cavities, limited to roughly $50$~MV/m, are therefore excluded, and
normal-conducting high-gradient structures become necessary.
At a higher accelerating gradient ($E_{acc}$), a higher acceleration efficiency,
  that is, a higher shunt impedance per unit length ($R_{sh}/L$), is required
 since the RF power per unit length ($P_{in}/L$) required to excite $E_{acc}$ is proportional to the square of $E_{acc}$
 according to:
\begin{equation}
  E_{acc} = \sqrt{\frac{R_{sh}}{L}\frac{P_{in}}{L}} ~. \label{eq:Eacc}
\end{equation}
The required $R_{sh}/L$ for the vertical linear collider can be evaluated based on the typical shunt impedance of the X-band accelerating structures
for the CLIC, that is, $R_{sh}/L$ $\approx$ $100$~M$\Omega$/m.
To excite a gradient of $E_{acc}$ $=$ $100$~MV/m, as in the baseline of the CLIC $3$-TeV version with that structure,
an RF power of $P_{in}/L$ $\approx$ $100$~MW/m is required, as obtained from Eq.~(\ref{eq:Eacc}).
If we excite a gradient of $E_{acc}$ $=$ $300$~MV/m for a $500$-GeV vertical linear collider with the above RF power, 
the required shunt impedance will be $R_{sh}/L$ $=$ $900$~M$\Omega$/m.
Therefore, a shunt impedance of $\approx 1$~G$\Omega$/m is a reasonable target for the vertical linear collider.
There are two current or near-term technologies that can be used to achieve such a high shunt impedance.
First, X-band dielectric-assist accelerating (DAA) structures \cite{daa:xband} offer a very high shunt impedance of
  $R_{sh}/L$ $\approx$ $500$~M$\Omega$/m at room temperature. This value
 can be dramatically increased to $1.8$~G$\Omega$/m based on the data in \cite{daa:cryo} if the DAA structure consists of five regular cells and two end cells per tube formed by pure magnesia and 6N copper, and the structure is cooled down to $77$~K \cite{daa:D.Satoh}.
It should be noted that breakdown as observed in metallic (all-metal) structures
  is strongly suppressed in DAA structures since fields on metallic surfaces are largely reduced and no current flows in dielectric materials below the dielectric breakdown threshold\footnote{The actual threshold for a short duration (i.e., shorter than 1~$\mu$s) in the microwave range is unknown; this research is being conducted at Nextef2 in KEK.}.
However, the gradient is presently limited by multipacting; this issue is currently under active investigation \cite{daa:xband}.
Second, higher acceleration frequencies offer both higher $E_{acc}$ and higher $R_{sh}/L$.
Novel normal-conducting metallic structures with a $300$-GHz acceleration frequency,
  which can be considered as the highest frequency of microwaves,
are coming within reach \cite{300GHz:Str},
  together with a $300$-GHz high-power source being developed at RIKEN \cite{300GHz:Src1,300GHz:Src2} for high-energy accelerators.
Since the shunt impedance increases as the acceleration frequency ($f_{acc}$) increases
   according to $R_{sh}/L$ $\propto$ $\sqrt{f_{acc}}$,
the expected shunt impedance of the $300$-GHz metallic structure is $R_{sh}/L$ $\approx$ $500$~M$\Omega$/m at room temperature.
Cooling the structure down to $77$~K increases the shunt impedance further, by a factor of
$\approx2$ at $300$~GHz, due to the improved electric conductivity of the cavity material (copper).
This factor is limited not only by material purity but also by the anomalous skin effect at $77$~K.
The expected shunt impedance is then $R_{sh}/L\approx1.0$~G$\Omega$/m, which still meets the target.
It is known that breakdown in metallic structures is significantly suppressed
  at cryogenic temperatures \cite{CryoXband,CryoDC}.

It should be noted that the two-beam power delivery scheme described in
Sec.~\ref{sec:example}\,\ref{sec:example:DriveBeam} is presently matched to the X-band option:
a CLIC-type combiner ring naturally provides drive-beam bunch spacings in the gigahertz range,
whereas driving a $300$-GHz structure would require harmonic power extraction from a
sub-harmonically bunched drive beam, with drive-bunch lengths well below
$\lambda/2\pi\approx0.16$~mm to maintain a usable form factor.
Power extraction structures at $300$~GHz have not been demonstrated; alternatively, the $300$-GHz option may rely on direct
 sub-terahertz sources \cite{300GHz:Src1,300GHz:Src2} distributed along the shaft.
%
The X-band cryogenic DAA structure is therefore regarded as the primary option,
 but the $300$-GHz cryogenic metallic structure is a prospective alternative
 if multipacting in DAA structures remains unresolved.
The above-mentioned technologies are highly challenging, but within reach.

\paragraph{Ultra-low-emittance high-current electron gun.}
The luminosity of a linear collider is set, to a large extent, at the source.
The transverse emittance delivered by the electron gun at the bottom of the shaft propagates
through the machine and, in the absence of an electron damping ring, largely determines the
achievable spot size at the IP.
Recent progress in cryogenic RF photoinjectors is therefore highly relevant.
Operating a copper photocathode gun at cryogenic temperature permits very high
peak surface fields and reduced intrinsic (thermal) emittance, yielding an
order-of-magnitude increase in five-dimensional beam brightness relative to that of
room-temperature guns~\cite{cryogun1,cryogun2}. Such ultra-low-emittance,
high-brightness sources are attractive for the vertical linear collider because they allow us to eliminate an electron damping ring
 and their cryogenic operating point is naturally
 compatible with the cryogenic main linac.
A spin-polarized variant of the
electron source could, in addition, provide possibilities
to open new channels for BSM searches as described in Sec.~\ref{sec:Concept:merits}.

\paragraph{Civil engineering: feasibility of $\sim$3-km vertical shaft.}
The vertical geometry converts the accelerator length into shaft depth,
so the feasibility of constructing a deep, large-diameter vertical shaft must be examined.
We adopt a working depth of $\sim$3~km.
This value is well matched to a Higgs factory machine length at the gradients discussed above,
and it sits at the frontier of, rather than beyond, demonstrated human-made
 underground excavation, which keeps the assumption reasonable
  while still spanning the required energy range.

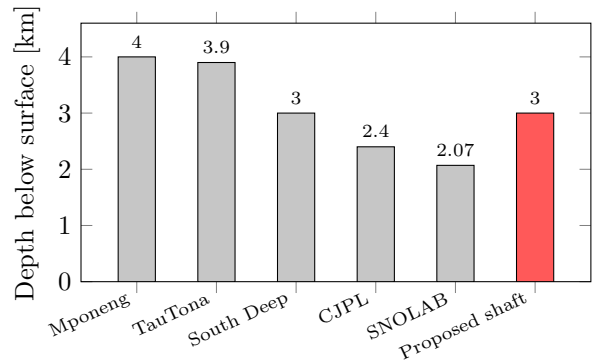
\begin{figure}[t]
\centering
\begin{tikzpicture}
\begin{axis}[
  width=\columnwidth, height=5.0cm,
  ybar, bar width=14pt,
  ymin=0, ymax=4.6,
  ylabel={Depth below surface [km]},
   symbolic x coords={Mponeng,TauTona,South~Deep,CJPL,SNOLAB,Proposed~shaft},
  xtick={Mponeng,TauTona,South~Deep,CJPL,SNOLAB,Proposed~shaft}, x tick label style={rotate=25,anchor=east,font=\scriptsize},
  ytick={0,1,2,3,4},
  enlarge x limits=0.14,
  nodes near coords, nodes near coords style={font=\scriptsize},
  every node near coord/.append style={anchor=south},
 bar shift=0pt,
]
\addplot[fill=gray!45] coordinates {(Mponeng,4.0) (TauTona,3.9) (South~Deep,3.0) (CJPL,2.4) (SNOLAB,2.07)};
\addplot[fill=red!65] coordinates {(Proposed~shaft,3.0)};
\end{axis}
\end{tikzpicture}
\caption{Approximate depths of representative deep human-made underground
excavations, namely ultra-deep South African gold mines (Mponeng, TauTona, South Deep)~\cite{durrheim,SouthDeep1,SouthDeep2},
China Jinping Underground Laboratory  (CJPL) \cite{CJPL}, and SNOLAB~\cite{snolab}, compared
with the $\sim$3-km vertical shaft assumed in this work (red). Values are approximate
and compiled from public sources.}
\label{fig:depth}
\end{figure}

The feasibility argument rests on publicly available precedent.
Ultra-deep hard-rock mining in the Witwatersrand Basin, South Africa, has reached depths approaching $4$~km, with workings and shaft infrastructure
operated routinely at $3$--$4$~km, including a hoisting shaft with a depth of $\sim$3~km sunk in
a single lift at the South Deep gold mine, which could be a precedent for the geometry assumed
here, and a century of accumulated rock engineering experience at these
depths~\cite{durrheim,durrheim2}.
Independently, the particle-physics community itself
operates large underground facilities at multi-kilometer depths; SNOLAB, for example, occupies excavated space $\sim$2.07~km below the surface
within an operating mine~\cite{snolab}. Figure~\ref{fig:depth} places the assumed
$3$-km shaft in the context of such existing excavations.

The principal engineering challenges at this depth, namely the geothermal gradient (rock temperature
rising by tens of degrees), the high lithostatic stress and associated rockburst
hazard, and groundwater inflow~\cite{durrheim,durrheim2}, are well characterized and can be
addressed with established countermeasures, such as forced ventilation and refrigeration,
heavy ground support, water control, and, importantly, staged construction in
which auxiliary inclined access and intermediate horizontal drifts are advanced
ahead of, and in parallel with, the main shaft. The difficulty is reduced
substantially at a geologically favorable site, that is, a site with a competent crystalline basement,
a moderate geothermal gradient, and the absence of active faults, a combination
that, as discussed in Sec.~\ref{sec:example}\,\ref{sec:example:site}, is plausibly met beneath the KEK Tsukuba
Campus. On this basis, we conclude that the excavation of a $\sim$3-km vertical
shaft is considered feasible with current or near-term technology, and that
adopting a $3$-km depth is a reasonable working assumption at the present
conceptual stage.

\subsection{Advantages\label{sec:Concept:merits}}
The vertical orientation offers several distinctive advantages beyond the compact
surface footprint already noted.

\emph{Intrinsic azimuthal symmetry.} With gravity acting along the beam axis,
the static mechanical load on linac components and their supports is,
ideally, azimuthally symmetric about that axis, rather than breaking the
transverse symmetry as in a horizontal machine. Moreover, no correction for the
curvature of the Earth is needed along the linac. Both features are favorable
for the high-precision alignment and ultra-low-emittance beam transport on which
luminosity depends.

\emph{Natural vertical alignment reference.} Alignment of a horizontal collider requires
an artificial straight reference (stretched wires, hydrostatic leveling networks, and an
explicit choice of how to treat the Earth's curvature) maintained over tens of kilometers.
In the vertical geometry, the reference line is provided by nature: the local plumb line
coincides with the beam axis, and a single vertical reference (an optical/laser plummet
combined with gyroscopic azimuth transfer, both established in deep-shaft surveying) can
span the entire machine.
Furthermore, each component can determine its own orientation locally, without an external
survey network, using precision tiltmeters that read the angle to the gravity vector
directly at the nrad level; together with the azimuthally symmetric static load noted
above, this favors the micron-level alignment on which emittance preservation depends.
A laser reference in a vertical shaft is also less affected by refraction, since the
atmospheric density gradient is parallel, rather than transverse, to the line of sight.
Residual issues, such as the azimuthal (roll) reference around the vertical axis,
ventilation-induced air path disturbances, and slow tilts of the local vertical due to
Earth tides, are real but small, and the latter fall well within the bandwidth of
beam-based feedback.
As a vertical-specific systematic, the Coriolis force of the Earth's rotation deflects the
upward- and downward-going beams in opposite east-west directions, by a sub-micrometer
displacement and nrad-level angle over the linac length at the latitude of Tsukuba;
this is a static, precisely calculable effect, removed by a small (and sign-asymmetric
between the two linacs) corrector setting, but it must be included in a future complete optics model.

\emph{Vibration isolation.} The IP and detector sit deep underground, where
mechanical noise generated on the surface is strongly attenuated;
this quiet environment is exploited by deep underground laboratories~\cite{CJPL,snolab}.
This is beneficial for the nano-beam stability required at the final focus.

\emph{Cosmically quiet interaction point.} At the IP depth of $\sim$1.5~km
($\approx 4$~km water equivalent in crystalline rock), 
the cosmic-ray muon flux may be reduced by approximately six orders of magnitude
 relative to that at the surface, 
to a level similar to that for deep underground laboratories of comparable overburden~\cite{snolab}.
For a collider detector, this is a qualitatively new environment. Searches for long-lived
particles, displaced vertices, and other detached or out-of-time signatures, which at
surface facilities are limited by or require elaborate vetoes against the cosmic
flux, can be conducted essentially cosmic-free.
The shaft itself offers the option of instrumenting sections above and below the
detector as low-background decay volumes along the boosted forward directions.
Beam-induced backgrounds remain and will dominate near the IP; the merit lies in
the near-total removal of the one background that no surface or shallow facility can shield.

\emph{Gravity-assisted cryogen transport.} If the main linac is cooled with liquid
nitrogen, the vertical geometry assists the cryogen cycle in both directions.
Since a continuous $3$-km liquid column is excluded by the critical pressure of nitrogen
($3.4$~MPa, corresponding to a $\sim$400-m head), the supply line is divided into
pressure-staged cells on the order of $100$--$200$~m, each with a small phase-separator
reservoir. A pump-free thermosiphon scheme of the kind is widely used in accelerator and
detector cryogenics.
Within each cell, the structures are cooled by natural convection; the downward supply from
reservoir to reservoir is driven by gravity itself. 
The warmed gas returns to the surface by buoyancy alone, 
so that all compressors, reliquefiers, and heat-rejection equipment stay above ground.
The vertical shaft thus eliminates the long horizontal cryogen-transfer and heat-rejection
lines of a conventional collider and confines active cryogenic machinery to the
surface. This is an operational and reliability benefit specific to this geometry, at the cost of
the staged-supply hardware, whose engineering design is left to a dedicated study.

\emph{Natural dump locations.} 
After passing the IP,
 the spent electron beam 
  is to be dumped at the surface with full maintenance access,
and the spent positron beam is to be dumped downward to a sealed dump at the shaft bottom shielded by the $3$-km overburden, as illustrated in Fig.~\ref{fig:complex}(a). 

\emph{Electricity self-sufficiency.} A large, deep, vertical infrastructure
creates an unusual opportunity to supply part of the facility's substantial
electrical load on site, thereby improving energy self-sufficiency and reducing both
operating cost and the carbon footprint.
Several mechanisms are enabled, in principle, by the excavation:
 the multi-kilometer vertical head is well suited to underground pumped-storage hydroelectricity
  and other gravity-based energy storage for load leveling;
  the elevated rock temperature at depth is a geothermal resource for heat and possibly low-grade power;
  and large underground caverns can host compressed air or other energy storage,
   as well as biomass/biogas conversion.
It should be noted that these possibilities are presented qualitatively,
  as an opportunity that merits dedicated study,
 and that
  no quantitative energy balance is claimed here.
The aspiration is that a future large accelerator facility could approach electricity self-sufficiency by
 co-developing its underground space as an energy asset.

\emph{Reuse of existing infrastructure.} Siting the machine at the KEK Tsukuba Campus allows
extensive reuse of the SuperKEKB complex~\cite{superkekb,superkekb2}, as detailed in Sec.~\ref{sec:example}.
 
\emph{New channels for BSM searches.}
No collider has ever been designed or operated with both its beam axis and the particle spins aligned with the local gravitational field.
While any direct gravitational effect on the collision kinematics is minute,
this configuration could acquire a meaning within the Standard Model Extension, 
for example
  the general effective field theory framework for Lorentz and CPT   
    violation~\cite{sme:cpt1,sme:cpt2},
  spin-gravity anomalous couplings~\cite{sme:grav},
  and axion-mediated monopole-dipole interactions sourced by the 
    Earth~\cite{moody:wilczek}.
The \epem\ configuration is, moreover, unique in that it collides a polarized particle moving
 against $\vec{g}$ with a polarized antiparticle moving along $\vec{g}$, realizing matter and antimatter spin-gravity orientations simultaneously.
We emphasize that stringent low-energy bounds (torsion pendula, comagnetometers) already
exist on many of these couplings~\cite{sme:tables}; whether a collider environment competitively probes
any combination is precisely the theoretical question the proposed concept
poses. 
 
\emph{Unique opportunity in social/industrial evaluation.}
The proposed concept conveys a strong message in terms of public relations, education, and international attention,
and offers a unique opportunity in terms of garnering public support and raising funds.
Furthermore, establishing technologies and know-how to construct, operate, utilize, and maintain such a large deep facility could give rise to a new industry
that generates enormous value.
The above points are extremely important for making a large project a reality.

\subsection{Disadvantages and Challenges\label{sec:Concept:demerits}}
The proposed concept has serious disadvantages and open challenges, several of
which are critical at the present stage.
\emph{(i) Accelerating gradient.} The required gradients of much higher than $100$~MV/m for linear colliders have
           not yet been demonstrated. Sustained research and development is needed to achieve such gradients (see Sec.~\ref{sec:Concept}\,\ref{sec:Concept:Tech}).
\emph{(ii) Civil engineering.} Excavating and utilizing a $\sim$3-km
shaft is a major, multi-year, high-cost undertaking with non-trivial
geothermal, geomechanical, and hydrological risks~\cite{durrheim,durrheim2}.
\emph{(iii) Installation and maintenance.} Installing, aligning, and maintaining accelerator components distributed along a deep
vertical shaft is intrinsically harder than doing so in a horizontal tunnel. In addition, hoisting capacity in a single vertical shaft imposes practical limits.
\emph{(iv) Luminosity and beam dynamics.} The luminosity, the compaction of the beam-delivery system, and the
beam-beam performance remain to be established using a start-to-end simulation.
\emph{(v) Physics motivation of the vertical geometry.} The notion that vertical,
gravity-aligned collisions with gravity-parallel polarization open a genuine BSM channel
remains to be established: it requires a dedicated analysis, for example within the Standard Model Extension framework (see Sec.~\ref{sec:Concept}\,\ref{sec:Concept:merits}).
Until such a study is conducted, this aspect should be regarded as a well-posed
theoretical question rather than a physics case. 
The cosmic-quiet merit of the deep IP, by
contrast, is not speculative; however, its exploitation requires a detector and trigger design
adapted to long-lived-particle signatures.

\section{Example of Accelerator Complex\label{sec:example}}
\begin{figure*}[t]
\centering
\includegraphics[width=.99\linewidth]{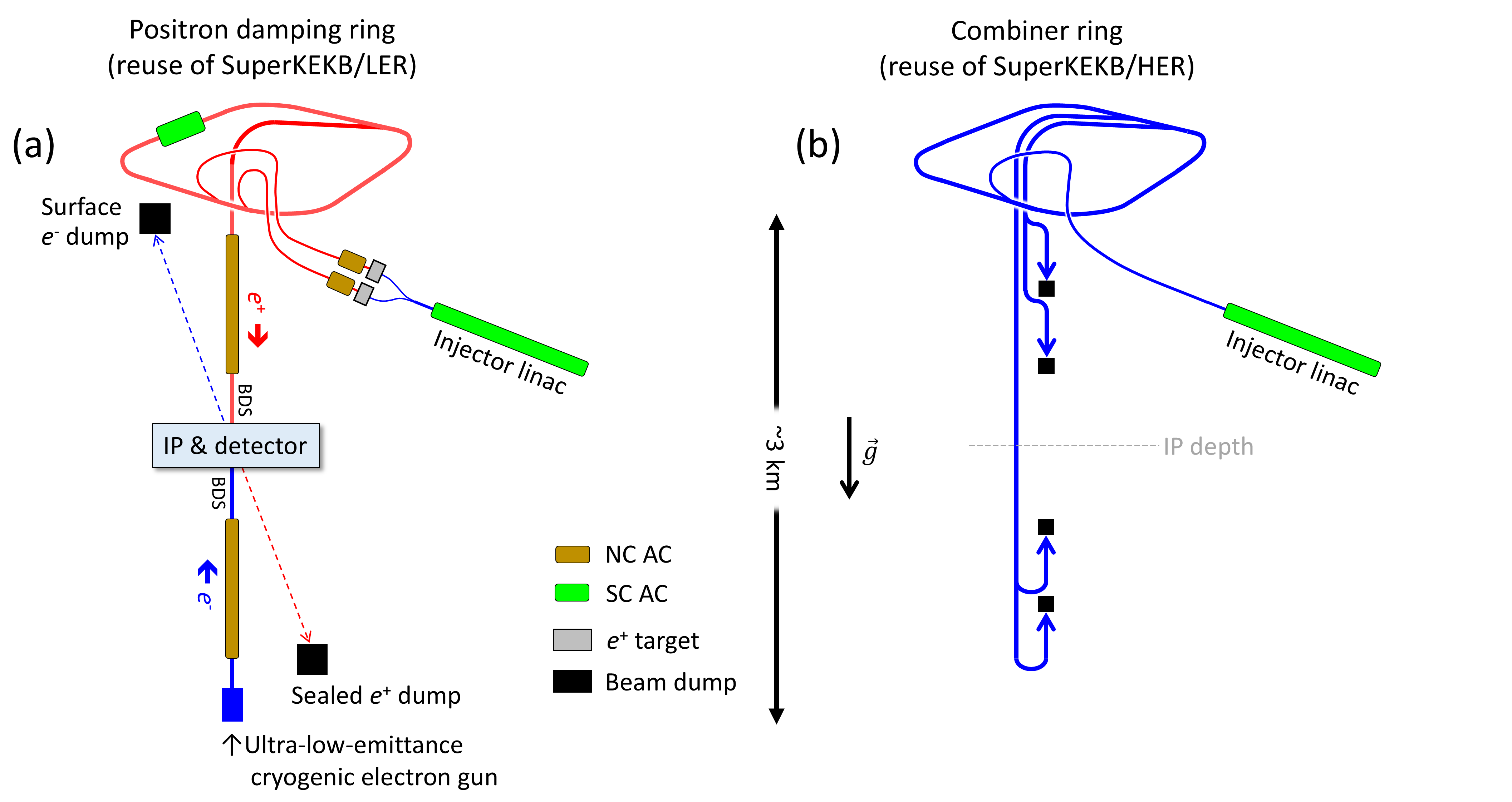}
  \caption{Example of accelerator complex at KEK Tsukuba Campus.
(a) Colliding \epem\ beams with sources and spent-beam dumps. The spent electron
beam is extracted to a dump at the surface, and the spent
positron beam is extracted to a sealed dump at the shaft bottom.
(b) Drive beams with combiner ring. The drive beams are transported downward from the
surface for both the main linacs,
 where only two sectors are drawn per main linac, and
the decelerator sections run alongside the main linacs shown in (a) within the same shaft.
Both rings are housed in the single circular tunnel presently used for SuperKEKB.
NC (SC) AC indicates a normal-conducting (super-conducting) accelerating structure.
The blue (red) elements indicate $e^-$ ($e^+$) beam lines.
Not to scale. LER: low-energy ring; HER: high-energy ring; BDS: beam-delivery system.
}
\label{fig:complex}
\end{figure*}

To clarify the proposed concept,
 we provide a concrete illustration
  of an accelerator complex that reuses the KEK Tsukuba Campus and the SuperKEKB facility~\cite{superkekb,superkekb2}.
This extensive reuse of SuperKEKB is central to the cost argument and to the choice of the Tsukuba site.
Figure~\ref{fig:complex} schematically shows the rough layout.

\subsection{Site of Vertical Shaft\label{sec:example:site}}
The KEK Tsukuba Campus is a natural candidate site. The site-specific case (see Sec.~\ref{sec:Concept}\,\ref{sec:Concept:Tech}) can be made concrete in three respects.
First, while the area around KEK near the surface is covered by relatively soft Quaternary strata, at a relatively shallow depth of around $400$~m below ground, a solid bedrock of granite, continuing from Mt. Tsukuba, lies firmly beneath the surface \cite{kanto,kanto2,tsukuba}. Due to this robust underground geological structure (crystalline bedrock), the ground in this area is extremely stable across a wide area.
Second, the Kanto region hosts one of the densest deep seismic and borehole
observation networks in the world, including instrumented boreholes with a depth of $2$--$3$~km, reaching the basement~\cite{nied},
so that the deep stress field, velocity structure, and ground-motion attenuation,
 which are design inputs for a deep shaft, can be quantitatively
 estimated. 
Third, downhole-surface observation pairs consistently show that ground motion within competent basement rock at depth is
substantially smaller than that at the surface, since deep structures are not subject to surface amplification.
The region is further characterized by a moderate geothermal gradient. The on-site
presence of accelerator infrastructure and expertise is another advantage.
A dedicated geotechnical site characterization (deep pilot boreholes, stress measurement,
and a hydrogeological survey) would be an early and essential step of any feasibility study.

\subsection{Colliding \epem\ Beams and Detector\label{sec:example:IP}}
The two main linacs are housed in the vertical shaft. Electrons from an
ultra-low-emittance high-current gun at the bottom are accelerated upward,
while positrons from the reused SuperKEKB low-energy ring (with some upgrades) as a damping ring
are accelerated downward. The beams collide at the IP near
mid-depth, where the detector is installed. 
A finite crossing angle at the IP should be generated in both or one of the beam-delivery systems.
The main linac uses high-gradient, high-efficiency,
cryogenic normal-conducting structures, cooled by liquid
nitrogen, as in \Cthree, with the gravity-assisted, pressure-staged cryogen transport
described in Sec.~\ref{sec:Concept}\,\ref{sec:Concept:merits}.
The deep, quiet IP is advantageous for the final focus and for the
detector. A compact beam-delivery system, integrating diagnostics, collimation, crossing-angle generation, 
and final focus, is assumed; its detailed design is left to future work.
The generation and preservation of positron polarization is also left to future work.

\emph{Spent-beam handling.}
The upward-going spent electron beam is extracted to a dump at the surface, $\sim$10~m laterally
offset from the shaft head assuming a crossing angle of $14$~mrad,
 where a conventional water dump with full maintenance access can
be used, while the downward-going spent positron beam is swept and absorbed in a sealed solid dump
 with closed-loop cooling at the shaft bottom.
The $3$-km overburden acts as an effectively infinite shield, so that the required containment grade is
substantially lighter than that for a surface dump, and the rejected heat, available at a usable
temperature level, can be fed into geothermal/energy-recovery systems (see Sec.~\ref{sec:Concept}\,\ref{sec:Concept:merits}).

\subsection{Drive Beam\label{sec:example:DriveBeam}}
To avoid distributing many high-power RF sources along a deep shaft, a two-beam
acceleration scheme of the type developed for the CLIC is adopted.
A high-current drive beam runs parallel to the main linac and
transfers its RF energy, via power extraction structures, to the accelerating
structures of the main beam. The SuperKEKB high-energy ring, a $7$-GeV
electron storage ring~\cite{superkekb}, can be reused (with some upgrades)
to generate and condition the drive beams, with an appropriate bunch structure
formed using an upgraded superconducting injector linac.
Here, reuse of the high-energy ring refers to the tunnel, magnets, and infrastructure. For the
transient bunch combination process, the ring is converted to isochronous optics with
injection/extraction RF deflectors operating at a sub-harmonic of the main-linac frequency
(e.g.,\ $2.856~\mathrm{GHz}=11.424~\mathrm{GHz}/4$, the KEK S-band)
 as in the CLIC combiner rings.
The required isochronous ($\alpha_p\approx0$) lattice and the control of coherent synchrotron radiation effects for the high-charge
trains require, however, a substantial optics redesign and belong to a dedicated study in the future.
Two drive beams are transported downward from the surface for both the main linacs,
 as shown in Fig.~\ref{fig:complex}(b).

\subsection{Injector Linac and Sources\label{sec:example:linac}}
The existing KEK injector linac, which currently fills the SuperKEKB
rings and the photon factory rings, is reused to supply the colliding-positron and drive-beam complexes
 using accelerating structures upgraded to superconducting ones for high-duty operation.
The SuperKEKB low-energy ring, a $4$-GeV positron storage ring,
 can be reused (with some upgrades) as the positron damping ring; here, reuse refers to the tunnel, magnets, and
infrastructure.
The positron-source chain doubling (or more) the chain enables us to operate it
 well below the damage threshold of the positron target.
An appropriate bunch structure for the acceleration frequency of the main linac
 is formed in the combiner ring (reuse of the high-energy ring), filled by the superconducting injector
linac in high-duty operation.
The electron beam for collisions is provided by an ultra-low-emittance high-current cryogenic electron gun,
  where relevant operational experience and expertise from the SuperKEKB low-emittance, high-current electron gun \cite{Yoshida:2014osa} may be utilized.

\section{Summary and Future Prospects}
We proposed a gravity-aligned vertical \epem\ linear collider as a concept
to address the cost-and-scale impasse facing conventional Higgs factories.
By orienting the beam axis along gravity and folding the machine into a single deep
vertical shaft, this concept offers a small surface footprint, intrinsic mechanical symmetry advantages, deep vibration isolation, a natural vertical alignment reference, gravity-assisted
(pressure-staged) cryogen transport, a cosmically quiet interaction point, natural
spent-beam dump locations, extensive reuse of the existing KEK/SuperKEKB complex,
an opportunity for on-site electricity self-sufficiency, and a unique opportunity in social/industrial evaluation.
Using only publicly available information, we argued that excavating a $\sim$3-km vertical shaft is feasible at a geologically favorable site such as the KEK Tsukuba Campus.
The proposed concept has the potential to break through the cost-and-scale impasse of conventional Higgs factories: (i) the acceleration technologies (see Sec.~\ref{sec:Concept}\,\ref{sec:Concept:Tech}) are challenging, but within reach since they are a mere extension of conventional microwave acceleration; 
(ii) we could expand the scope of funding outside the academic field based on the expected unique opportunity in social/industrial evaluation.

The proposed concept, which was presented here only at the conceptual level, has several disadvantages. The required accelerating gradient, the cost and risk of deep-shaft
construction, the difficulty of installation and maintenance at depth, and the unproven luminosity all require dedicated study.
The most valuable next steps are a demonstration of high-gradient cryogenic DAA structures and $300$-GHz structures; 
a geotechnical site study and shaft-construction plan; 
the development of cryogenic ultra-low-emittance high-current polarized electron sources;
a start-to-end luminosity simulation that includes a compact beam-delivery system and
vertical-specific systematics in beam optics, such as the Coriolis deflection and the tidal tilt of the local vertical; 
the generation and preservation of positron polarization; 
a theoretical investigation, formulated within the Standard Model Extension
and related frameworks for spin-gravity
couplings~\cite{sme:cpt1,sme:cpt2,sme:grav,sme:tables,moody:wilczek}, of whether vertical,
gravity-aligned collisions with gravity-parallel polarization constrain any coefficient
combinations beyond existing bounds; 
a study of the long-lived-particle physics program
enabled by the cosmically quiet deep interaction point;
and the development of a set of strawman machine parameters.
%
The goal of this paper is for this concept to inspire such studies across the
high-energy, particle-physics, and astrophysics communities.


\end{document}